\documentclass{article}
\usepackage{rotating}

\begin{document}

\title{Error-Prone Cellular Automata as Metaphors of Immunity as Computation}

\author{K\'atia~K.~Cassiano\\
Valmir~C.~Barbosa\thanks{Corresponding author (valmir@cos.ufrj.br).}\\
\\
Programa de Engenharia de Sistemas e Computa\c c\~ao, COPPE\\
Universidade Federal do Rio de Janeiro\\
Caixa Postal 68511\\
21941-972 Rio de Janeiro - RJ, Brazil}

\date{}

\maketitle

\begin{abstract}
Modeling the immune system so that its essential functionalities stand out
without the need for every molecular or cellular interaction to be taken into
account has been challenging for many decades. Two competing approaches have
been the clonal selection theory and the idiotypic-network theory, each stemming
from a relatively separate set of principles. One recent perspective holding the
promise of unification is that of immunity as computation, that is, of immunity
as the process of computing the state of the body so that protection can be
effected, as well as boosted through learning. Here we investigate the use of
cellular automata (CA) as the core abstraction supporting this new perspective.
Our choice of CA for this role is based on the potential variety of basins in
a CA attractor field. Associating each basin with a consistent set of body
states, and moreover providing for the noisy evolution of the CA in time so that
jumping between basins is possible, have provided the necessary backdrop. Given
a CA rule to be followed by all cells synchronously, our model is based on a
probability with which each cell, at each time step, independently updates its
own state differently than the rule mandates. Setting up and solving the
corresponding Markov chain for its stationary probabilities have revealed that
already in the context of elementary CA there exist rules that, while allowing
transitions between basins, display remarkable resiliency in terms of basin
occupation. For these rules, the long-run probability that the CA is found in a
given basin is practically the same as in the deterministic case when the
initial CA state is chosen uniformly at random. We argue that, consequently, our
single-parameter CA model may be a suitable abstraction of immunity as
computation.

\end{abstract}

\newpage
\section{Introduction}\label{sec:intr}

The immune system is one of the body's major regulatory systems. Comprising
important elements at various physical scales, such as organs, cells, and
molecules, the immune system provides defenses against pathogenic bacteria and
viruses, identifies and seeks to eliminate abnormally behaving cells before
they become established tumors, and carries out tissue restoration as well as
various other housekeeping activities. The immune response to invading
pathogens, as well as the system's participation in body maintenance, are the
product of learning and self-organization: Beginning with the so-called innate
immunity, the immune system is capable of recreating itself along its history
while avoiding the pitfalls of autoimmunity \cite{nci}. In order to remain fit
for such potentially daunting task for as long as possible, the immune system
relies on the process known as somatic hypermutation \cite{os06}, which
continually provides the required diversity at the immune-cellular level.

While by virtue of the immune system's nature as a self-organizing entity it
seems safe to view the rise of the various immune functions as a process that
proceeds from the bottom up, starting with local interactions at the molecular
level, immunity is undoubtedly a systemic process. Explanatory theories of the
immune system have therefore oscillated between the very local (with the clonal
selection theory \cite{b59,f95}) and the very wide (with the elusive
idiotypic-network theory \cite{j74,b07,mkbmqtcb11}, based on the idea that many
immune-system elements interact with one another much as they do with antigens).
A curious (though apt) perspective that might reconcile the two extremes is that
the immune system continually ``computes'' the state of the body (of which it is
part), resulting in state alterations as the immune system both acts and learns
\cite{c07}.

Models of the immune system, however, have concentrated on expressing the
evolution in time of cell concentrations and other quantities, usually by
differential equations (cf., e.g., \cite{fabc04}) but also by discrete-time
abstractions akin to cellular automata (CA) \cite{stb12}. In general such models
have been shown to provide a qualitatively convincing picture of how several of
the important immune functions arise, or of how the idiotypic network is thought
to be organized. But the immunity-as-computation paradigm is to our knowledge
yet to be explored, though it should be for at least two reasons that we find
quite compelling. The first one is that viewing immunity as resulting from the
continual computation of states of the body is bound to require new abstractions
through which such states can be represented and manipulated, mathematically or
computationally. As a consequence, valuable insight can be expected to emerge.
The second reason is that, once suitable state representations have been
identified, the possibility of uncertain events that renders the entire system
both adaptive and vulnerable can be more easily taken into account.

Here we begin to investigate the use of CA as a suitable abstraction to underlie
the study of the immune system as a computational entity. Although choosing CA
may seem only natural to unconditional CA enthusiasts, given the impressive
plethora of domains to which CA have been applied \cite{w02}, in our vision
there are specific reasons backing our choice. One of them is that, by virtue of
the deterministic character of how CA evolve in time, all CA states for a given
finite number of cells and a fixed rule are necessarily partitioned into
attractor basins. Viewing CA states as body states and the CA rule as
summarizing the computation of body states by the immune system immediately
yields an interpretation of each basin as the set of states to which the body is
confined once it is born into that basin. Depending on the CA rule in question
some basins may express a complex succession of body states while others may
seem dull by comparison or merely bespeak decay and disorganization.

Another reason for choosing a representation by CA is that they yield easily to
the incorporation of uncertainty. This can be achieved in many ways, our choice
being to allow each cell, at each time step, to disobey the CA rule in use and
change its state differently than the rule mandates. We model this possibility
by a single probability parameter, denoted by $p$. The usual, deterministic CA
world is recovered by setting $p$ to $0$, but proceeding otherwise (i.e.,
choosing $p>0$) immediately opens up new doors. Specifically, every CA state
becomes reachable from every other state, whence it follows that the
aforementioned attractor basins are no longer unreachable from one another
during the CA dynamics but rather allow the body whose states are the CA states
to journey through a rich variety of domains (health, disease, recovery, etc.),
however unlikely the transition from one to another may be. It also follows that
the attractor dynamics inside a basin is no longer inevitable, and likewise that
the periodic attractor lying at a basin's core is not inescapable.

The question we seek to answer is the following. Given a CA rule and an
attractor basin in the corresponding CA-state space, what is the probability
that, in the long run, the CA state is part of that basin? Unlike other studies
that models uncertainty in a manner similar to ours (cf., e.g., \cite{sc13} and
references therein), answering this question relies not on analyzing
spatiotemporal patterns of CA evolution but rather on solving Markov
chains for their stationary distributions. This is computationally strenuous,
but for modestly sized systems we show that there do exist CA rules for which
the added uncertainty, while allowing the desired transitions between CA states
of different basins to occur, nevertheless tends to confine the CA dynamics to
within the same basin where it would unfold if no uncertainty had been added
but initial conditions were random.

We proceed in the following manner. We present our model, along with its main
properties, in Section~\ref{sec:model}. This is followed by our methodology in
Section~\ref{sec:methods}, results in Section~\ref{sec:results}, and discussion
in Sections~\ref{sec:disc1} and~\ref{sec:disc2}. We conclude in
Section~\ref{sec:concl}.

\section{Model}\label{sec:model}

We consider binary CA, i.e., CA whose cell states are either $0$ or $1$. If $n$
is the number of cells, assumed finite, then the number of distinct CA states is
$2^n$. All cells update their states at all times synchronously (i.e., in
lockstep) based on the same rule, which can be thought of as a table of binary
outputs indexed by $(\delta+1)$-bit inputs. Here $\delta$ is the size of a
cell's neighborhood, the same for all cells, so a cell's new state depends on
its own current state and on its neighbors' current states. Each rule's size is
$2^{\delta+1}$, so there exist $2^{2^{\delta+1}}$ distinct rules. Fixing the
rule to be used gives rise to a function $f$ mapping each CA state in
$\{0,1\}^n$ into another state in the same set.

Our model is based on turning deterministic CA into probabilistic ones. We do
this by introducing a probability, $p$, with which each cell, at each time step,
disobeys the rule's prescription for its next state independently of all other
cells. So, if $x$ denotes a cell's next state and the CA rule's current
prescription for the value of $x$ is $b\in\{0,1\}$, we have
\begin{equation}
x:=\left\{
\begin{array}{ll}
1-b,&\mbox{with probability $p$;}\\
b,&\mbox{otherwise.}
\end{array}
\right.
\label{eq:disobey}
\end{equation}

Now let $i,j\in \{0,1\}^n$ be any two CA states and let $H_{i,j}$ be the
Hamming distance between them (i.e., the number of cells at which $i$ and $j$
differ). Additionally, let $k_i=f(i)$, i.e., $k_i$ is the CA state that follows
$i$ in the deterministic dynamics for the rule at hand. Once we introduce the
probability $p$, the probability that CA state $i$ is followed by $j$, denoted
by $p_{i,j}$, is
\begin{equation}
p_{i,j}=
p^{H_{j,k_i}}(1-p)^{n-H_{j,k_i}}.
\label{eq:pij}
\end{equation}
Readily, letting $j=k_i$ yields $H_{j,k_i}=0$, and consequently
$p_{i,j}=(1-p)^n$. This is the probability with which $i$ is followed by $k_i$,
that is, the probability that at any given time step the deterministic
prescription is respected.

Thus, while using $p=0$ clearly recovers the traditional, deterministic
dynamics (since $p_{i,j}=1$ if $j=k_i$ and $p_{i,j}=0$ otherwise), using $p>0$
lets the CA dynamics be described as a discrete-time Markov chain on the CA
states and having $P=[p_{i,j}]$ for transition-probability matrix. To see this
it suffices to verify that the elements of $P$ sum up to $1$ on any row. That
is, fixing $i$ yields
\begin{equation}
\sum_{j\in \{0,1\}^n}p_{i,j}=
\sum_{h=0}^n{n\choose h}p^h(1-p)^{n-h}=1
\end{equation}
(because $k_i$ is fixed along with $i$ and differs at $h$ cells from
${n\choose h}$ of the $2^n$ CA states for any given number $h$ of cells).
Moreover, for $p>0$ every element of $P$ is nonzero and therefore the chain is
ergodic, meaning that, regardless of how likely it is for any given CA state to
be the initial state, in the long run the CA is found in state $i$ with the
stationary probability $\pi_i$ given by $\pi=\pi P$, where $\pi=[\pi_i]$ is a
row vector.

\subsection{On symmetry}

By Eq.~(\ref{eq:disobey}), letting $p=1$ also implies deterministic behavior,
but following the rule that is complementary to the one that is followed when
$p=0$. That is, one rule sets $x$ to $b$ if and only if the other sets it to
$1-b$. A similar type of symmetry occurs between the case in which $p>0$ and
that in which $1-p$ is used instead.

To see this, first let $\bar l$ denote the complement of CA state $l$ (i.e.,
adding any cell's state in $l$ to its state in $\bar l$ yields $1$). It clearly
follows that $H_{l,j}+H_{\bar l,j}=n$ for any CA state $j$. Now recall that
Eq.~(\ref{eq:pij}) refers to a specific CA rule and to each cell disobeying it
with probability $p$ at each time step. Rewriting the equation for the
complementary rule and also letting it be disobeyed with probability $1-p$
instead has no effect on the value of $p_{i,j}$, since
\begin{equation}
(1-p)^{H_{j,\bar k_i}}p^{n-H_{j,\bar k_i}}=
(1-p)^{n-H_{j,k_i}}p^{H_{j,k_i}}.
\end{equation}
Thus, studying the case of any given rule under $p$ leads to the same Markov
chain as studying the complementary rule under $1-p$, and consequently to the
same stationary probabilities on the CA states.

Typically our interest lies in small values of $p$, which makes the case of (the
correspondingly large) $1-p$ even more remarkable, at least at the level of CA
states. At the higher level of the attractor basins, however, no equivalence can
in general be expected: The probability that the CA is found in a particular
basin in the long run depends on how the CA states cluster into basins and in
general this happens differently for a given rule and its complement.

Nevertheless, there do exist rule pairs that display equivalent behavior for the
same value of $p$. We identify these pairs by first introducing a transformation
between CA states, call it $g$, and requiring that one of the rules in the pair
lead the CA from state $i$ to state $k_i$ if and only if the other rule leads
the CA from state $g(i)$ to state $g(k_i)$. Any rule pair satisfying this
requirement is such that the corresponding sets of attractor basins, one for
each rule, are structurally equivalent to each other. If, moreover, we require
$H_{j,k_i}=H_{g(j),g(k_i)}$, then we also have $p_{i,j}=p_{g(i),g(j)}$. What
results from this is that, in the long run, the CA is found in any given basin
of one of the rules with the same probability that it is found in the equivalent
basin of the other rule.

Rule pairs like this are important in our context because they have the
potential of reducing the number of rules that have to be analyzed. This is so
because, even though the two sets of stationary probabilities on the CA states
are in general distinct, when the probabilities are summed up inside any basin
of one of the rules the result is the same as that for the other rule's
equivalent basin. One transformation $g$ for which every rule has a counterpart
with which it satisfies the two requirements above is negation, i.e., adding any
cell's state in $i$ to its state in $g(i)$ yields $1$. Another one is
reflection, i.e., the $c$th cell's state in $i$ is the same as the $(n-c+1)$th
cell's state in $g(i)$ for every $c\in\{1,2,\ldots,n\}$.

\subsection{A special case}

By Eq.~(\ref{eq:pij}), letting $p=0.5$ leads to $p_{i,j}=1/2^n$ regardless of
$i$, $j$, or the rule being used. From this it follows that $\pi_i=1/2^n$ for
every $i$, so the CA is equally likely to be found at any state in the long run.
However, our transition-probability matrix $P$ for this particular value of $p$
is not the only one leading to the uniform distribution over the CA states: In
fact, this happens if and only if the matrix is doubly stochastic (i.e., its
elements add up to $1$ column-wise just as they do row-wise) and implies an
ergodic chain. An example is obtained by letting
\begin{equation}
p_{i,j}=\left\{
\begin{array}{ll}
1/{n\choose\tau},&\mbox{if $H_{i,j}=\tau$;}\\
0,&\mbox{otherwise}
\end{array}
\right.
\end{equation}
for any number $\tau$ of cells \cite{vb04} (but note that our $p=0.5$ case is
not equivalent to choosing any particular value for $\tau$).

\subsection{The general case}

Our model is a special case of the so-called probabilistic CA (PCA), in which
a cell's next state is no longer given by the customary deterministic rule but
instead is chosen probabilistically as a function of the cell's and its
neighbors' current states. Our particular type of PCA relies on the
probabilistic decision summarized in Eq.~(\ref{eq:disobey}), itself dependent on
a specific deterministic rule (unlike most PCA, in whose case no deterministic
rule plays any role).

Placing our model within the wider class of PCA is important because they have
been viewed as prototypes of many important systems, both physical and
computational, in a way similar to that in which immunity may come to be
characterized as a computational process. Examples of such systems include the
spin lattices of statistical physics \cite{dk84,bg85,gjh85,gl89,lms90} and, more
generally, the Markov and Gibbs random fields \cite{b93} that, together with
various asynchronous state-update schemes \cite{ib84,bg89}, underlie many of the
so-called probabilistic graphical models (such as Bayesian networks and hidden
Markov models) in artificial intelligence \cite{kf09}.

\section{Methods}\label{sec:methods}

Given a deterministic CA rule and the number $n$ of cells, let $m$ denote the
number of attractor basins into which the set $\{0,1\}^n$ is partitioned. We
denote these basins by $B_1,B_2,\ldots,B_m$. For the case in which the rule in
question may be disobeyed by any cell at any time step according to
Eq.~(\ref{eq:disobey}) with $p>0$, our aim is to calculate the probability that,
in the long run, the CA is found in some state of a given basin
$B\in\{B_1,B_2,\ldots,B_m\}$. Denoting this probability by $\pi_B$, we clearly
have
\begin{equation}
\pi_B=\sum_{k_0\in \{0,1\}^n}\pi_{B\mid k_0}\mathrm{Pr}(k_0),
\end{equation}
where $\pi_{B\mid k_0}$ is the conditional probability that in the long run the
CA is found in some state in $B$, given that it started at state $k_0$, and
$\mathrm{Pr}(k_0)$ is the probability that it did start at $k_0$. However, it
follows from our discussion in Section~\ref{sec:model} that $\pi_{B\mid k_0}$ is
actually unaffected by $k_0$ and can be obtained by adding up $\pi_k$, the
stationary probability of CA state $k$ in the associated Markov chain, for all
$k\in B$. We then have
\begin{equation}
\pi_B=\sum_{k\in B}\pi_k,
\end{equation}
regardless of how we choose the initial state $k_0$, i.e., regardless of
$\mathrm{Pr}(k_0)$ for any $k_0$.

All our analyses in the forthcoming sections are based on comparing $\pi_B$ to
the corresponding probability when $p=0$, that is, when evolution is
deterministic. We denote this probability by $\sigma_B$ and the corresponding
conditional probability, given $k_0$, by $\sigma_{B\mid k_0}$. Readily,
\begin{equation}
\sigma_{B\mid k_0}=\left\{
\begin{array}{ll}
1,&\mbox{if $k_0\in B$;}\\
0,&\mbox{otherwise}
\end{array}
\right.
\end{equation}
and
\begin{equation}
\sigma_B=
\sum_{k_0\in \{0,1\}^n}\sigma_{B\mid k_0}\mathrm{Pr}(k_0)=
\sum_{k_0\in B}\mathrm{Pr}(k_0),
\end{equation}
so $\sigma_B$ is clearly dependent upon how $k_0$ is chosen. We continue by
assuming that this happens uniformly at random, that is,
$\mathrm{Pr}(k_0)=1/2^n$ for every $k_0$, whence we obtain
\begin{equation}
\sigma_B=\frac{\vert B\vert}{2^n}.
\end{equation}
Thus, $\sigma_B$ results trivially from the uniform distribution over all CA
states (we simply add it up for all states in basin $B$).

Obtaining $\pi_B$ for every basin $B$ requires the system $\pi=\pi P$ to be
solved, subject to the constraints that $\pi_i>0$ for all $i\in \{0,1\}^n$ and
$\sum_{i\in \{0,1\}^n}\pi_i=1$, for each desired combination of $n$, CA rule,
and $p>0$. We have used the solver that is freely available as part of the
Tangram-II modeling tool \cite{slsrdfjm06}. This solver employs state-of-the-art
techniques for the above determination of $\pi$ given $P$, but in our case
$P$ is a $2^n\times 2^n$ matrix with no zeroes and no facilitating symmetries or
structure. Thus the solution process has been very time-consuming, which has
constrained $n$ to the modest values of $10$ through $12$. For the record, we
mention that, depending on the CA rule at hand, stepping up to $n=13$ would
demand nearly two months per run on an Intel Xeon E5-1650 at 3.2GHz with enough
memory to store the entire $8192\times 8192$ system at all times. This,
unfortunately, has proven infeasible.

\section{Results}\label{sec:results}

Henceforth we let $\mathcal{B}$ denote the set $\{B_1,B_2,\ldots,B_m\}$ of all
basins for a given CA rule and a fixed value of $n$. We compare the
distributions $\pi_{B_1},\pi_{B_2},\ldots,\pi_{B_m}$ and
$\sigma_{B_1},\sigma_{B_2},\ldots,\sigma_{B_m}$ by means of the Hellinger
distance between them, denoted by $H(\pi,\sigma)$ and given by
\begin{equation}
H(\pi,\sigma)=\sqrt{1-\sum_{B\in\mathcal{B}}\sqrt{\pi_B\sigma_B}}.
\end{equation}
Using the Hellinger distance to compare the two distributions is convenient not
only because it truly is a distance function but also because it is always such
that $0\le H(\pi,\sigma)\le 1$. In fact, clearly $H(\pi,\sigma)=0$ if and only
if $\pi_B=\sigma_B$ for all $B\in\mathcal{B}$ and $H(\pi,\sigma)=1$ if and only
if $\pi_B\sigma_B=0$ for all $B\in\mathcal{B}$. The latter, however, can never
be achieved in our context because both $\pi_B$ and $\sigma_B$ are strictly
positive for all $B\in\mathcal{B}$.

We also compare the mean and standard deviation of basin sizes as they vary from
one distribution to the other. To this end, we use the ratios
\begin{equation}
\rho_\mathrm{mean}=
\frac
{\sum_{B\in\mathcal{B}}\pi_B\vert B\vert}
{\sum_{B\in\mathcal{B}}\sigma_B\vert B\vert}
\end{equation}
and
\begin{equation}
\rho_\mathrm{s.d.}=
\sqrt{\frac
{\sum_{B\in\mathcal{B}}\pi_B\vert B\vert^2
-(\sum_{B\in\mathcal{B}}\pi_B\vert B\vert)^2}
{\sum_{B\in\mathcal{B}}\sigma_B\vert B\vert^2
-(\sum_{B\in\mathcal{B}}\sigma_B\vert B\vert)^2}
}.
\end{equation}
Clearly, comparing $\rho_\mathrm{mean}$ to $1$ lets us detect increases or
decreases in the mean basin size as we move from using the probabilities
$\sigma_{B_1},\sigma_{B_2},\ldots,\sigma_{B_m}$ to using
$\pi_{B_1},\pi_{B_2},\ldots,\pi_{B_m}$, and likewise for $\rho_\mathrm{s.d.}$
with respect to the standard deviation of basin sizes.

These data are given in Tables~\ref{table:dist} and~\ref{table:msd}, the former
containing Hellinger distances, the latter containing mean and
standard-deviation ratios. All data refer to elementary CA \cite{w83}, which in
the present context corresponds to setting a cell's neighborhood size ($\delta$)
to $2$, and to an arrangement of cells that is one-dimensional with periodic
boundaries (i.e., the first and last cells in the arrangement are neighbors).
Moreover, our data encompass all combinations of a unique rule, a CA size
$n\in\{10,11,12\}$, and a probability $p\in\{0.001,0.01\}$. By unique rule we
mean one that is not equivalent to any other selected rule by negation or
reflection. Of the $256$ possible elementary-CA rules, $88$ are unique in this
sense but group with the remaining $168$ rules into equivalence classes of size
at most $4$, or into larger clusters of size at most $8$ as two equivalence
classes of pairwise complementary rules are joined. Each of the equivalence
classes might be represented in our tables by any of its members, but we follow
Wuensche and Lesser, who in their atlas \cite{wl92} use one or two rules of each
larger cluster, viz.\ the rule of least number (in the customary Wolfram sense
\cite{w83}) and its complement if not already in the first rule's equivalence
class. Each table also informs a rule's class (1 through 4) according to
Wolfram's well-known qualitative scheme \cite{w84}.

\begin{sidewaystable}[p]
\centering
\caption{Hellinger distances.}
\label{table:dist}
\begin{tabular}{|c|c|cc|cc|cc|}
\hline
&Wolfram	&\multicolumn{2}{|c|}{$H(\pi,\sigma)$ for $n=10$}	&\multicolumn{2}{|c|}{$H(\pi,\sigma)$ for $n=11$}	&\multicolumn{2}{|c|}{$H(\pi,\sigma)$ for $n=12$}\\
\cline{3-8}
Rule	&class	&$p=0.001$	&$p=0.01$	&$p=0.001$	&$p=0.01$	&$p=0.001$	&$p=0.01$\\
\hline
\bf 0	&\bf 1	&\bf 0.000000	&\bf 0.000000	&\bf 0.000000	&\bf 0.000000	&\bf 0.000000	&\bf 0.000000\\
248	&1	&0.248956	&0.248918	&0.223216	&0.223206	&0.200278	&0.200274\\
\bf 249	&\bf 1	&\bf 0.091299	&\bf 0.091276	&\bf 0.073387	&\bf 0.073380	&\bf 0.059551	&\bf 0.059549\\
250	&1	&0.176776	&0.176729	&0.015626	&0.015626	&0.125000	&0.124995\\
\bf 251	&\bf 1	&\bf 0.031257	&\bf 0.031224	&\bf 0.000000	&\bf 0.000000	&\bf 0.015626	&\bf 0.015623\\
\bf 252	&\bf 1	&\bf 0.022100	&\bf 0.022100	&\bf 0.015626	&\bf 0.015626	&\bf 0.011049	&\bf 0.011049\\
\bf 253	&\bf 1	&\bf 0.000000	&\bf 0.000000	&\bf 0.000000	&\bf 0.000000	&\bf 0.000000	&\bf 0.000000\\
\bf 254	&\bf 1	&\bf 0.022100	&\bf 0.022100	&\bf 0.015626	&\bf 0.015626	&\bf 0.011049	&\bf 0.011049\\
\hline
1	&2	&0.237315	&0.239354	&0.260833	&0.261584	&0.283045	&0.283760\\
2	&2	&0.214709	&0.203996	&0.224895	&0.223983	&0.234612	&0.233671\\
3	&2	&0.121415	&0.124573	&0.136260	&0.136566	&0.150518	&0.150766\\
4	&2	&0.198416	&0.178198	&0.207861	&0.206196	&0.216913	&0.215180\\
5	&2	&0.122062	&0.123784	&0.136260	&0.136566	&0.151784	&0.152007\\
6	&2	&0.145932	&0.099881	&0.153279	&0.140815	&0.178978	&0.167294\\
7	&2	&0.570014	&0.073262	&0.600195	&0.477705	&0.627823	&0.494088\\
9	&2	&0.127448	&0.058579	&0.198415	&0.169627	&0.130059	&0.112595\\
10	&2	&0.104076	&0.102039	&0.088408	&0.086590	&0.106435	&0.104504\\
11	&2	&0.339538	&0.265903	&0.290900	&0.238953	&0.543483	&0.346141\\
\bf 12	&\bf 2	&\bf 0.084915	&\bf 0.083186	&\bf 0.088408	&\bf 0.086590	&\bf 0.091966	&\bf 0.090065\\
13	&2	&0.561884	&0.436485	&0.298189	&0.248901	&0.620558	&0.469093\\
14	&2	&0.296465	&0.245507	&0.335487	&0.266519	&0.492403	&0.338750\\
\bf 15	&\bf 2	&\bf 0.000000	&\bf 0.000000	&\bf 0.000000	&\bf 0.000000	&\bf 0.000000	&\bf 0.000000\\
\hline
\end{tabular}

\end{sidewaystable}

\addtocounter{table}{-1}
\begin{sidewaystable}[p]
\centering
\caption{Continued.}
\begin{tabular}{|c|c|cc|cc|cc|}
\hline
&Wolfram	&\multicolumn{2}{|c|}{$H(\pi,\sigma)$ for $n=10$}	&\multicolumn{2}{|c|}{$H(\pi,\sigma)$ for $n=11$}	&\multicolumn{2}{|c|}{$H(\pi,\sigma)$ for $n=12$}\\
\cline{3-8}
Rule	&class	&$p=0.001$	&$p=0.01$	&$p=0.001$	&$p=0.01$	&$p=0.001$	&$p=0.01$\\
\hline
\it 19	&\it 2	&\it 0.665262	&\it 0.455025	&\it 0.687454	&\it 0.466017	&\it 0.706507	&\it 0.476426\\
\it 23	&\it 2	&\it 0.649786	&\it 0.504885	&\it 0.674805	&\it 0.517551	&\it 0.699909	&\it 0.532909\\
24	&2	&0.151170	&0.145988	&0.162003	&0.156480	&0.163455	&0.157906\\
25	&2	&0.178794	&0.142287	&0.205302	&0.166376	&0.240752	&0.184615\\
\bf 26	&\bf 2	&\bf 0.096811	&\bf 0.087081	&\bf 0.092313	&\bf 0.081963	&\bf 0.082902	&\bf 0.073380\\
\bf 27	&\bf 2	&\bf 0.078410	&\bf 0.075206	&\bf 0.078149	&\bf 0.075674	&\bf 0.088139	&\bf 0.084074\\
28	&2	&0.507050	&0.293820	&0.276705	&0.187250	&0.554002	&0.304611\\
\bf 29	&\bf 2	&\bf 0.042041	&\bf 0.041289	&\bf 0.044401	&\bf 0.043618	&\bf 0.046283	&\bf 0.045463\\
33	&2	&0.128619	&0.124749	&0.102505	&0.099605	&0.131497	&0.128095\\
35	&2	&0.112197	&0.095522	&0.108425	&0.093326	&0.136032	&0.100919\\
36	&2	&0.210118	&0.202473	&0.212746	&0.204591	&0.218307	&0.209832\\
37	&2	&0.217930	&0.128618	&0.130832	&0.085161	&0.153067	&0.127335\\
\bf 38	&\bf 2	&\bf 0.053281	&\bf 0.051025	&\bf 0.060564	&\bf 0.058097	&\bf 0.058462	&\bf 0.055931\\
41	&2	&0.120070	&0.097433	&0.115620	&0.106492	&0.150240	&0.107285\\
43	&2	&0.299150	&0.254258	&0.340329	&0.279313	&0.499819	&0.357524\\
46	&2	&0.176524	&0.167568	&0.186924	&0.177534	&0.196353	&0.186419\\
50	&2	&0.621971	&0.428828	&0.434170	&0.334403	&0.667557	&0.448039\\
\bf 51	&\bf 2	&\bf 0.000000	&\bf 0.000000	&\bf 0.000000	&\bf 0.000000	&\bf 0.000000	&\bf 0.000000\\
57	&2	&0.271948	&0.093908	&0.095220	&0.034859	&0.287305	&0.079804\\
58	&2	&0.230614	&0.194367	&0.401245	&0.276011	&0.410814	&0.410734\\
62	&2	&0.197845	&0.138003	&0.132255	&0.122913	&0.215650	&0.117917\\
77	&2	&0.649786	&0.504885	&0.447864	&0.375344	&0.699909	&0.532909\\
\hline
\end{tabular}

\end{sidewaystable}

\addtocounter{table}{-1}
\begin{sidewaystable}[p]
\centering
\caption{Continued.}
\begin{tabular}{|c|c|cc|cc|cc|}
\hline
&Wolfram	&\multicolumn{2}{|c|}{$H(\pi,\sigma)$ for $n=10$}	&\multicolumn{2}{|c|}{$H(\pi,\sigma)$ for $n=11$}	&\multicolumn{2}{|c|}{$H(\pi,\sigma)$ for $n=12$}\\
\cline{3-8}
Rule	&class	&$p=0.001$	&$p=0.01$	&$p=0.001$	&$p=0.01$	&$p=0.001$	&$p=0.01$\\
\hline
94	&2	&0.306439	&0.269040	&0.277744	&0.242535	&0.569790	&0.280718\\
178	&2	&0.649786	&0.504885	&0.447864	&0.375344	&0.699909	&0.532909\\
197	&2	&0.529675	&0.345739	&0.285589	&0.211139	&0.581408	&0.362983\\
198	&2	&0.522388	&0.330289	&0.282790	&0.203589	&0.572138	&0.344295\\
201	&2	&0.220405	&0.209932	&0.217887	&0.207934	&0.221112	&0.209812\\
\bf 204	&\bf 2	&\bf 0.000000	&\bf 0.000000	&\bf 0.000000	&\bf 0.000000	&\bf 0.000000	&\bf 0.000000\\
\bf 205	&\bf 2	&\bf 0.081927	&\bf 0.080898	&\bf 0.082447	&\bf 0.081349	&\bf 0.083984	&\bf 0.082815\\
\bf 210	&\bf 2	&\bf 0.007328	&\bf 0.006725	&\bf 0.000000	&\bf 0.000000	&\bf 0.010027	&\bf 0.008364\\
212	&2	&0.299150	&0.254258	&0.340329	&0.279313	&0.499819	&0.357524\\
214	&2	&0.104722	&0.100715	&0.138109	&0.122696	&0.112897	&0.094736\\
217	&2	&0.120698	&0.115389	&0.130762	&0.125581	&0.126279	&0.121099\\
218	&2	&0.266122	&0.259242	&0.221638	&0.214302	&0.265582	&0.258562\\
\bf 220	&\bf 2	&\bf 0.092624	&\bf 0.090070	&\bf 0.092633	&\bf 0.089882	&\bf 0.094236	&\bf 0.091322\\
\bf 222	&\bf 2	&\bf 0.081912	&\bf 0.081363	&\bf 0.081185	&\bf 0.080555	&\bf 0.084533	&\bf 0.084083\\
226	&2	&0.170992	&0.148976	&0.079509	&0.075465	&0.196928	&0.168250\\
227	&2	&0.128235	&0.089781	&0.076123	&0.065573	&0.109403	&0.072279\\
228	&2	&0.322168	&0.308522	&0.308795	&0.298360	&0.299346	&0.290927\\
229	&2	&0.101086	&0.096509	&0.134254	&0.119041	&0.115874	&0.108862\\
230	&2	&0.238324	&0.230492	&0.258104	&0.250046	&0.277262	&0.268729\\
\it 232	&\it 2	&\it 0.649786	&\it 0.504885	&\it 0.674805	&\it 0.517551	&\it 0.699909	&\it 0.532909\\
233	&2	&0.370172	&0.311163	&0.404103	&0.341928	&0.420107	&0.357435\\
\it 236	&\it 2	&\it 0.767929	&\it 0.677742	&\it 0.790184	&\it 0.701010	&\it 0.810056	&\it 0.722260\\
\hline
\end{tabular}

\end{sidewaystable}

\addtocounter{table}{-1}
\begin{sidewaystable}[p]
\centering
\caption{Continued.}
\begin{tabular}{|c|c|cc|cc|cc|}
\hline
&Wolfram	&\multicolumn{2}{|c|}{$H(\pi,\sigma)$ for $n=10$}	&\multicolumn{2}{|c|}{$H(\pi,\sigma)$ for $n=11$}	&\multicolumn{2}{|c|}{$H(\pi,\sigma)$ for $n=12$}\\
\cline{3-8}
Rule	&class	&$p=0.001$	&$p=0.01$	&$p=0.001$	&$p=0.01$	&$p=0.001$	&$p=0.01$\\
\hline
237	&2	&0.511025	&0.446457	&0.532280	&0.465832	&0.551887	&0.483901\\
\bf 240	&\bf 2	&\bf 0.000000	&\bf 0.000000	&\bf 0.000000	&\bf 0.000000	&\bf 0.000000	&\bf 0.000000\\
241	&2	&0.161638	&0.159585	&0.169371	&0.167225	&0.176764	&0.174527\\
242	&2	&0.219813	&0.216880	&0.229854	&0.226784	&0.239672	&0.236473\\
\bf 243	&\bf 2	&\bf 0.084915	&\bf 0.083186	&\bf 0.088408	&\bf 0.086590	&\bf 0.091966	&\bf 0.090065\\
244	&2	&0.253371	&0.248050	&0.265043	&0.259493	&0.276340	&0.270564\\
246	&2	&0.123588	&0.121820	&0.126632	&0.124803	&0.133269	&0.131387\\
\hline
18	&3	&0.177219	&0.171971	&0.098671	&0.095609	&0.216715	&0.200709\\
22	&3	&0.051738	&0.044548	&0.090170	&0.077895	&0.250064	&0.134211\\
\bf 30	&\bf 3	&\bf 0.034544	&\bf 0.016889	&\bf 0.020255	&\bf 0.007665	&\bf 0.038162	&\bf 0.009492\\
\bf 45	&\bf 3	&\bf 0.016456	&\bf 0.010594	&\bf 0.000000	&\bf 0.000000	&\bf 0.056506	&\bf 0.004097\\
\bf 60	&\bf 3	&\bf 0.000000	&\bf 0.000000	&\bf 0.000000	&\bf 0.000000	&\bf 0.000000	&\bf 0.000000\\
73	&3	&0.294481	&0.182849	&0.120781	&0.107784	&0.181444	&0.151175\\
\bf 90	&\bf 3	&\bf 0.000000	&\bf 0.000000	&\bf 0.000000	&\bf 0.000000	&\bf 0.000000	&\bf 0.000000\\
\bf 105	&\bf 3	&\bf 0.000000	&\bf 0.000000	&\bf 0.000000	&\bf 0.000000	&\bf 0.000000	&\bf 0.000000\\
126	&3	&0.114009	&0.110473	&0.158805	&0.146309	&0.140149	&0.125633\\
\bf 150	&\bf 3	&\bf 0.000000	&\bf 0.000000	&\bf 0.000000	&\bf 0.000000	&\bf 0.000000	&\bf 0.000000\\
161	&3	&0.213897	&0.142957	&0.046676	&0.018741	&0.145264	&0.101305\\
182	&3	&0.118028	&0.098982	&0.037891	&0.030647	&0.110355	&0.095368\\
225	&3	&0.034380	&0.020135	&0.062034	&0.018674	&0.488660	&0.119769\\
\hline
54	&4	&0.250242	&0.160801	&0.078368	&0.060045	&0.285840	&0.138826\\
\bf 193	&\bf 4	&\bf 0.066972	&\bf 0.049254	&\bf 0.057118	&\bf 0.030831	&\bf 0.096156	&\bf 0.052231\\
\hline
\end{tabular}

\end{sidewaystable}

\begin{sidewaystable}[p]
\centering
\caption{Mean and standard-deviation ratios (I: $p=0.001$; II: $p=0.01$).}
\label{table:msd}
\begin{tabular}{|c|c|cc|cc|cc|cc|cc|cc|}
\hline
&       &\multicolumn{4}{|c|}{$n=10$}   &\multicolumn{4}{|c|}{$n=11$}   &\multicolumn{4}{|c|}{$n=12$}\\
\cline{3-14}
&Wolfram        &\multicolumn{2}{|c|}{$\rho_\mathrm{mean}$}     &\multicolumn{2}{|c|}{$\rho_\mathrm{s.d.}$}     &\multicolumn{2}{|c|}{$\rho_\mathrm{mean}$}     &\multicolumn{2}{|c|}{$\rho_\mathrm{s.d.}$}     &\multicolumn{2}{|c|}{$\rho_\mathrm{mean}$}     &\multicolumn{2}{|c|}{$\rho_\mathrm{s.d.}$}\\
\cline{3-14}
Rule    &class  &I      &II     &I      &II     &I      &II     &I      &II     &I      &II     &I      &II\\
\hline
\bf 0	&\bf 1	&\bf 1.00	&\bf 1.00	&\bf 1.00	&\bf 1.00	&\bf 1.00	&\bf 1.00	&\bf 1.00	&\bf 1.00	&\bf 1.00	&\bf 1.00	&\bf 1.00	&\bf 1.00\\
248	&1	&1.13	&1.13	&0.00	&0.00	&1.11	&1.11	&0.00	&0.00	&1.09	&1.09	&0.00	&0.00\\
\bf 249	&\bf 1	&\bf 1.02	&\bf 1.02	&\bf 0.00	&\bf 0.00	&\bf 1.01	&\bf 1.01	&\bf 0.00	&\bf 0.00	&\bf 1.01	&\bf 1.01	&\bf 0.00	&\bf 0.00\\
250	&1	&1.06	&1.06	&0.00	&0.00	&1.00	&1.00	&0.00	&0.00	&1.03	&1.03	&0.00	&0.00\\
\bf 251	&\bf 1	&\bf 1.00	&\bf 1.00	&\bf 0.00	&\bf 0.00	&\bf 1.00	&\bf 1.00	&\bf 1.00	&\bf 1.00	&\bf 1.00	&\bf 1.00	&\bf 0.00	&\bf 0.00\\
\bf 252	&\bf 1	&\bf 1.00	&\bf 1.00	&\bf 0.00	&\bf 0.00	&\bf 1.00	&\bf 1.00	&\bf 0.00	&\bf 0.00	&\bf 1.00	&\bf 1.00	&\bf 0.00	&\bf 0.00\\
\bf 253	&\bf 1	&\bf 1.00	&\bf 1.00	&\bf 1.00	&\bf 1.00	&\bf 1.00	&\bf 1.00	&\bf 1.00	&\bf 1.00	&\bf 1.00	&\bf 1.00	&\bf 1.00	&\bf 1.00\\
\bf 254	&\bf 1	&\bf 1.00	&\bf 1.00	&\bf 0.00	&\bf 0.00	&\bf 1.00	&\bf 1.00	&\bf 0.00	&\bf 0.00	&\bf 1.00	&\bf 1.00	&\bf 0.00	&\bf 0.00\\
\hline
1	&2	&0.41	&0.40	&0.74	&0.74	&0.35	&0.35	&0.68	&0.68	&0.30	&0.30	&0.62	&0.62\\
2	&2	&0.75	&0.80	&0.98	&0.99	&0.72	&0.73	&0.93	&0.93	&0.67	&0.67	&0.88	&0.88\\
3	&2	&0.91	&0.88	&0.93	&0.94	&0.85	&0.85	&0.90	&0.90	&0.80	&0.80	&0.86	&0.85\\
4	&2	&0.49	&0.53	&0.72	&0.75	&0.45	&0.46	&0.68	&0.68	&0.42	&0.43	&0.64	&0.65\\
5	&2	&0.70	&0.65	&0.76	&0.71	&0.63	&0.63	&0.71	&0.70	&0.59	&0.58	&0.66	&0.65\\
6	&2	&0.88	&0.95	&1.05	&1.03	&0.85	&0.87	&1.12	&1.12	&0.84	&0.85	&1.02	&1.02\\
7	&2	&1.40	&1.07	&0.14	&1.00	&1.49	&1.44	&0.15	&0.50	&1.43	&1.38	&0.16	&0.51\\
9	&2	&1.11	&1.04	&0.97	&1.00	&0.84	&0.86	&1.58	&1.53	&0.99	&0.98	&0.98	&0.99\\
10	&2	&0.90	&0.90	&0.98	&0.98	&0.92	&0.92	&1.02	&1.02	&0.90	&0.90	&1.04	&1.04\\
11	&2	&1.08	&1.08	&0.88	&0.90	&1.11	&1.09	&0.90	&0.92	&2.37	&1.92	&0.39	&0.91\\
\bf 12	&\bf 2	&\bf 0.93	&\bf 0.93	&\bf 1.01	&\bf 1.01	&\bf 0.92	&\bf 0.92	&\bf 1.01	&\bf 1.01	&\bf 0.91	&\bf 0.92	&\bf 1.00	&\bf 1.00\\
13	&2	&1.93	&1.83	&0.18	&0.54	&1.17	&1.16	&0.08	&0.26	&2.33	&2.14	&0.24	&0.71\\
14	&2	&1.03	&1.03	&0.99	&0.99	&1.42	&1.37	&0.71	&0.77	&1.91	&1.71	&0.26	&0.68\\
\bf 15	&\bf 2	&\bf 1.00	&\bf 1.00	&\bf 1.00	&\bf 1.00	&\bf 1.00	&\bf 1.00	&\bf 1.00	&\bf 1.00	&\bf 1.00	&\bf 1.00	&\bf 1.00	&\bf 1.00\\
\hline
\end{tabular}

\end{sidewaystable}

\addtocounter{table}{-1}
\begin{sidewaystable}[p]
\centering
\caption{Continued.}
\begin{tabular}{|c|c|cc|cc|cc|cc|cc|cc|}
\hline
&       &\multicolumn{4}{|c|}{$n=10$}   &\multicolumn{4}{|c|}{$n=11$}   &\multicolumn{4}{|c|}{$n=12$}\\
\cline{3-14}
&Wolfram        &\multicolumn{2}{|c|}{$\rho_\mathrm{mean}$}     &\multicolumn{2}{|c|}{$\rho_\mathrm{s.d.}$}     &\multicolumn{2}{|c|}{$\rho_\mathrm{mean}$}     &\multicolumn{2}{|c|}{$\rho_\mathrm{s.d.}$}     &\multicolumn{2}{|c|}{$\rho_\mathrm{mean}$}     &\multicolumn{2}{|c|}{$\rho_\mathrm{s.d.}$}\\
\cline{3-14}
Rule    &class  &I      &II     &I      &II     &I      &II     &I      &II     &I      &II     &I      &II\\
\hline
\it 19	&\it 2	&\it 3.68	&\it 2.95	&\it 0.48	&\it 1.16	&\it 4.28	&\it 3.27	&\it 0.58	&\it 1.33	&\it 4.97	&\it 3.61	&\it 0.70	&\it 1.51\\
\it 23	&\it 2	&\it 3.33	&\it 3.01	&\it 0.26	&\it 0.75	&\it 3.89	&\it 3.43	&\it 0.32	&\it 0.88	&\it 4.57	&\it 3.92	&\it 0.38	&\it 1.02\\
24	&2	&1.01	&1.01	&1.23	&1.21	&0.97	&0.97	&1.03	&1.02	&0.97	&0.98	&0.98	&0.98\\
25	&2	&1.00	&0.99	&0.99	&1.00	&0.87	&0.89	&1.27	&1.22	&1.21	&1.17	&0.99	&1.01\\
\bf 26	&\bf 2	&\bf 0.92	&\bf 0.92	&\bf 1.03	&\bf 1.03	&\bf 0.94	&\bf 0.94	&\bf 0.98	&\bf 0.98	&\bf 0.89	&\bf 0.90	&\bf 0.94	&\bf 0.94\\
\bf 27	&\bf 2	&\bf 0.95	&\bf 0.95	&\bf 0.97	&\bf 0.97	&\bf 0.93	&\bf 0.93	&\bf 0.94	&\bf 0.94	&\bf 0.94	&\bf 0.94	&\bf 0.98	&\bf 0.98\\
28	&2	&2.16	&1.78	&0.35	&0.85	&1.22	&1.18	&0.16	&0.48	&2.72	&2.04	&0.46	&1.03\\
\bf 29	&\bf 2	&\bf 1.03	&\bf 1.03	&\bf 1.03	&\bf 1.03	&\bf 1.04	&\bf 1.04	&\bf 1.06	&\bf 1.06	&\bf 1.04	&\bf 1.04	&\bf 1.05	&\bf 1.05\\
33	&2	&0.77	&0.77	&0.85	&0.86	&0.79	&0.79	&0.84	&0.84	&0.77	&0.77	&0.83	&0.83\\
35	&2	&1.15	&1.13	&1.01	&1.01	&1.00	&1.00	&0.97	&0.97	&1.10	&1.07	&1.00	&1.00\\
36	&2	&0.57	&0.59	&0.90	&0.90	&0.54	&0.56	&0.86	&0.87	&0.50	&0.52	&0.82	&0.83\\
37	&2	&0.86	&0.92	&1.18	&1.10	&0.96	&0.99	&1.14	&1.09	&1.03	&1.03	&0.96	&0.96\\
\bf 38	&\bf 2	&\bf 0.96	&\bf 0.97	&\bf 1.02	&\bf 1.02	&\bf 0.96	&\bf 0.96	&\bf 1.04	&\bf 1.03	&\bf 0.95	&\bf 0.95	&\bf 1.00	&\bf 1.00\\
41	&2	&1.04	&1.03	&0.95	&0.97	&1.16	&1.15	&1.12	&1.12	&0.83	&0.89	&0.97	&1.01\\
43	&2	&1.06	&1.05	&0.86	&0.87	&1.42	&1.38	&0.68	&0.74	&2.08	&1.87	&0.24	&0.65\\
46	&2	&0.96	&0.96	&1.09	&1.08	&0.97	&0.97	&1.14	&1.13	&0.92	&0.93	&1.03	&1.03\\
50	&2	&3.29	&2.74	&0.37	&0.94	&1.58	&1.52	&0.14	&0.43	&4.49	&3.43	&0.52	&1.22\\
\bf 51	&\bf 2	&\bf 1.00	&\bf 1.00	&\bf 1.00	&\bf 1.00	&\bf 1.00	&\bf 1.00	&\bf 1.00	&\bf 1.00	&\bf 1.00	&\bf 1.00	&\bf 1.00	&\bf 1.00\\
57	&2	&1.28	&1.12	&0.34	&0.83	&1.03	&1.01	&0.28	&0.78	&1.35	&1.12	&0.41	&0.89\\
58	&2	&0.97	&0.98	&1.00	&0.99	&1.36	&1.30	&0.18	&0.52	&1.35	&1.31	&0.43	&0.43\\
62	&2	&1.24	&1.17	&0.61	&0.77	&1.31	&1.30	&1.17	&1.16	&0.85	&0.96	&1.04	&1.01\\
77	&2	&3.33	&3.01	&0.26	&0.75	&1.58	&1.55	&0.10	&0.32	&4.57	&3.92	&0.38	&1.02\\
\hline
\end{tabular}

\end{sidewaystable}

\addtocounter{table}{-1}
\begin{sidewaystable}[p]
\centering
\caption{Continued.}
\begin{tabular}{|c|c|cc|cc|cc|cc|cc|cc|}
\hline
&       &\multicolumn{4}{|c|}{$n=10$}   &\multicolumn{4}{|c|}{$n=11$}   &\multicolumn{4}{|c|}{$n=12$}\\
\cline{3-14}
&Wolfram        &\multicolumn{2}{|c|}{$\rho_\mathrm{mean}$}     &\multicolumn{2}{|c|}{$\rho_\mathrm{s.d.}$}     &\multicolumn{2}{|c|}{$\rho_\mathrm{mean}$}     &\multicolumn{2}{|c|}{$\rho_\mathrm{s.d.}$}     &\multicolumn{2}{|c|}{$\rho_\mathrm{mean}$}     &\multicolumn{2}{|c|}{$\rho_\mathrm{s.d.}$}\\
\cline{3-14}
Rule    &class  &I      &II     &I      &II     &I      &II     &I      &II     &I      &II     &I      &II\\
\hline
94	&2	&0.84	&0.87	&1.13	&1.12	&0.90	&0.92	&1.04	&1.03	&2.04	&1.30	&0.66	&1.08\\
178	&2	&3.33	&3.01	&0.26	&0.75	&1.58	&1.55	&0.10	&0.32	&4.57	&3.92	&0.38	&1.02\\
197	&2	&1.91	&1.70	&0.28	&0.75	&1.17	&1.15	&0.13	&0.39	&2.30	&1.92	&0.37	&0.94\\
198	&2	&2.17	&1.85	&0.30	&0.80	&1.22	&1.19	&0.14	&0.42	&2.75	&2.17	&0.41	&0.98\\
201	&2	&0.91	&0.91	&1.16	&1.15	&0.90	&0.91	&1.12	&1.11	&0.97	&0.96	&1.22	&1.19\\
\bf 204	&\bf 2	&\bf 1.00	&\bf 1.00	&\bf 1.00	&\bf 1.00	&\bf 1.00	&\bf 1.00	&\bf 1.00	&\bf 1.00	&\bf 1.00	&\bf 1.00	&\bf 1.00	&\bf 1.00\\
\bf 205	&\bf 2	&\bf 0.97	&\bf 0.97	&\bf 1.02	&\bf 1.01	&\bf 0.97	&\bf 0.97	&\bf 1.00	&\bf 1.00	&\bf 0.97	&\bf 0.97	&\bf 1.01	&\bf 1.01\\
\bf 210	&\bf 2	&\bf 1.00	&\bf 1.00	&\bf 1.00	&\bf 1.00	&\bf 1.00	&\bf 1.00	&\bf 1.00	&\bf 1.00	&\bf 1.00	&\bf 1.00	&\bf 1.00	&\bf 1.00\\
212	&2	&1.04	&1.04	&0.91	&0.91	&1.56	&1.50	&0.85	&0.89	&2.08	&1.87	&0.24	&0.65\\
214	&2	&0.96	&0.96	&1.05	&1.05	&0.79	&0.82	&1.09	&1.09	&0.97	&0.97	&0.96	&0.97\\
217	&2	&0.94	&0.94	&1.37	&1.36	&0.92	&0.92	&1.37	&1.35	&0.90	&0.90	&1.12	&1.11\\
218	&2	&0.59	&0.60	&0.80	&0.81	&0.55	&0.56	&0.79	&0.80	&0.50	&0.52	&0.73	&0.74\\
\bf 220	&\bf 2	&\bf 0.93	&\bf 0.93	&\bf 1.02	&\bf 1.02	&\bf 0.92	&\bf 0.92	&\bf 1.01	&\bf 1.01	&\bf 0.91	&\bf 0.92	&\bf 1.01	&\bf 1.01\\
\bf 222	&\bf 2	&\bf 0.85	&\bf 0.85	&\bf 0.91	&\bf 0.91	&\bf 0.83	&\bf 0.84	&\bf 0.89	&\bf 0.89	&\bf 0.82	&\bf 0.81	&\bf 0.88	&\bf 0.88\\
226	&2	&1.08	&1.07	&0.96	&0.97	&0.99	&0.99	&1.00	&1.00	&1.09	&1.08	&0.98	&0.98\\
227	&2	&1.07	&1.04	&1.00	&1.00	&1.07	&1.05	&1.04	&1.04	&1.08	&1.05	&1.02	&1.01\\
228	&2	&0.72	&0.74	&1.05	&1.05	&0.69	&0.70	&0.98	&0.98	&0.66	&0.68	&0.91	&0.91\\
229	&2	&0.96	&0.96	&1.03	&1.03	&0.89	&0.90	&1.01	&1.01	&0.80	&0.81	&0.86	&0.86\\
230	&2	&0.70	&0.71	&0.86	&0.87	&0.73	&0.74	&0.90	&0.91	&0.70	&0.71	&0.80	&0.81\\
\it 232	&\it 2	&\it 3.33	&\it 3.01	&\it 0.26	&\it 0.75	&\it 3.89	&\it 3.43	&\it 0.32	&\it 0.88	&\it 4.57	&\it 3.92	&\it 0.38	&\it 1.02\\
233	&2	&1.39	&1.37	&0.09	&0.28	&1.48	&1.46	&0.09	&0.28	&1.53	&1.51	&0.09	&0.29\\
\it 236	&\it 2	&\it 5.40	&\it 5.11	&\it 0.24	&\it 0.72	&\it 6.39	&\it 6.01	&\it 0.27	&\it 0.82	&\it 7.57	&\it 7.08	&\it 0.31	&\it 0.92\\
\hline
\end{tabular}

\end{sidewaystable}

\addtocounter{table}{-1}
\begin{sidewaystable}[p]
\centering
\caption{Continued.}
\begin{tabular}{|c|c|cc|cc|cc|cc|cc|cc|}
\hline
&       &\multicolumn{4}{|c|}{$n=10$}   &\multicolumn{4}{|c|}{$n=11$}   &\multicolumn{4}{|c|}{$n=12$}\\
\cline{3-14}
&Wolfram        &\multicolumn{2}{|c|}{$\rho_\mathrm{mean}$}     &\multicolumn{2}{|c|}{$\rho_\mathrm{s.d.}$}     &\multicolumn{2}{|c|}{$\rho_\mathrm{mean}$}     &\multicolumn{2}{|c|}{$\rho_\mathrm{s.d.}$}     &\multicolumn{2}{|c|}{$\rho_\mathrm{mean}$}     &\multicolumn{2}{|c|}{$\rho_\mathrm{s.d.}$}\\
\cline{3-14}
Rule    &class  &I      &II     &I      &II     &I      &II     &I      &II     &I      &II     &I      &II\\
\hline
237	&2	&1.88	&1.84	&0.10	&0.31	&2.00	&1.95	&0.10	&0.32	&2.13	&2.07	&0.11	&0.33\\
\bf 240	&\bf 2	&\bf 1.00	&\bf 1.00	&\bf 1.00	&\bf 1.00	&\bf 1.00	&\bf 1.00	&\bf 1.00	&\bf 1.00	&\bf 1.00	&\bf 1.00	&\bf 1.00	&\bf 1.00\\
241	&2	&0.80	&0.80	&0.94	&0.94	&0.77	&0.77	&0.90	&0.90	&0.76	&0.76	&0.88	&0.89\\
242	&2	&0.78	&0.78	&0.96	&0.96	&0.68	&0.68	&0.81	&0.81	&0.69	&0.69	&0.86	&0.86\\
\bf 243	&\bf 2	&\bf 0.94	&\bf 0.94	&\bf 1.04	&\bf 1.04	&\bf 0.92	&\bf 0.92	&\bf 1.02	&\bf 1.02	&\bf 0.92	&\bf 0.93	&\bf 1.00	&\bf 1.00\\
244	&2	&0.96	&0.96	&0.98	&0.98	&0.96	&0.96	&0.99	&0.99	&0.96	&0.96	&1.00	&1.00\\
246	&2	&0.88	&0.88	&1.10	&1.10	&0.89	&0.90	&1.17	&1.16	&0.92	&0.92	&1.07	&1.07\\
\hline
18	&3	&0.86	&0.86	&0.94	&0.94	&0.90	&0.90	&1.04	&1.04	&0.81	&0.82	&0.64	&0.65\\
22	&3	&0.94	&0.95	&0.98	&1.00	&0.97	&0.96	&0.99	&0.98	&0.60	&0.79	&1.01	&1.05\\
\bf 30	&\bf 3	&\bf 0.98	&\bf 0.99	&\bf 1.05	&\bf 1.04	&\bf 0.98	&\bf 0.99	&\bf 1.02	&\bf 1.01	&\bf 0.99	&\bf 1.00	&\bf 1.10	&\bf 1.04\\
\bf 45	&\bf 3	&\bf 0.99	&\bf 0.99	&\bf 1.01	&\bf 1.01	&\bf 1.00	&\bf 1.00	&\bf 1.00	&\bf 1.00	&\bf 0.98	&\bf 1.00	&\bf 1.00	&\bf 1.00\\
\bf 60	&\bf 3	&\bf 1.00	&\bf 1.00	&\bf 1.00	&\bf 1.00	&\bf 1.00	&\bf 1.00	&\bf 1.00	&\bf 1.00	&\bf 1.00	&\bf 1.00	&\bf 1.00	&\bf 1.00\\
73	&3	&1.08	&1.02	&1.16	&1.09	&0.93	&0.93	&0.90	&0.90	&0.91	&0.86	&0.98	&0.92\\
\bf 90	&\bf 3	&\bf 1.00	&\bf 1.00	&\bf 1.00	&\bf 1.00	&\bf 1.00	&\bf 1.00	&\bf 1.00	&\bf 1.00	&\bf 1.00	&\bf 1.00	&\bf 1.00	&\bf 1.00\\
\bf 105	&\bf 3	&\bf 1.00	&\bf 1.00	&\bf 1.00	&\bf 1.00	&\bf 1.00	&\bf 1.00	&\bf 1.00	&\bf 1.00	&\bf 1.00	&\bf 1.00	&\bf 1.00	&\bf 1.00\\
126	&3	&0.91	&0.91	&1.07	&1.07	&0.88	&0.88	&1.02	&1.02	&0.87	&0.88	&0.75	&0.78\\
\bf 150	&\bf 3	&\bf 1.00	&\bf 1.00	&\bf 1.00	&\bf 1.00	&\bf 1.00	&\bf 1.00	&\bf 1.00	&\bf 1.00	&\bf 1.00	&\bf 1.00	&\bf 1.00	&\bf 1.00\\
161	&3	&0.97	&0.98	&1.16	&1.12	&1.02	&1.00	&0.98	&1.02	&0.93	&0.94	&1.05	&1.05\\
182	&3	&1.02	&1.02	&0.87	&0.88	&1.02	&1.02	&0.95	&0.96	&1.01	&1.00	&0.87	&0.86\\
225	&3	&0.99	&0.99	&1.02	&1.02	&1.03	&0.99	&0.88	&1.04	&0.39	&0.92	&1.24	&1.14\\
\hline
54	&4	&0.86	&0.90	&0.84	&0.89	&1.03	&1.04	&1.06	&1.05	&1.41	&1.17	&0.86	&0.98\\
\bf 193	&\bf 4	&\bf 1.03	&\bf 1.02	&\bf 0.99	&\bf 1.00	&\bf 1.07	&\bf 1.03	&\bf 0.92	&\bf 0.97	&\bf 0.95	&\bf 0.97	&\bf 1.05	&\bf 1.03\\
\hline
\end{tabular}

\end{sidewaystable}

\section{Discussion}\label{sec:disc1}

As we have seen, disobeying a CA rule independently at each cell with
probability $p$ makes the CA dynamics stochastic and puts it between two
extremes that, in a sense, are equivalent. One extreme is the $p=0$ case, i.e.,
the case in which the rule is not disobeyed at all and the customary
deterministic dynamics is followed. In this case, the long-run probability that
a randomly chosen CA state is in some basin $B$ is $\sigma_B$ and stems from the
uniform distribution on the CA states provided the initial state is itself
chosen uniformly at random. The other extreme is that of $p=0.5$, in which case
the long-run probability that the CA is found in basin $B$ is $\pi_B$, now
stemming from CA-state probabilities that are again uniform but now by virtue of
the underlying Markov chain's stationary distribution.

Comparing these two distributions as indicated in Section~\ref{sec:results}
clearly yields $H(\pi,\sigma)=0$ and, consequently,
$\rho_\mathrm{mean}=\rho_\mathrm{s.d.}=1$, regardless of the particular CA rule
and CA size being considered. Although these values may look like what we seek
(stochastic CA dynamics that, while allowing occasional transitions between
basins, let the CA state be found in a same basin for long stretches of time),
they are only the product of erratic transitions between the CA states. In fact,
for $p=0.5$ all CA states are equally likely candidates for where the CA is to
move next, regardless of where it is currently.

It is instructive to contrast this $p=0.5$ extreme with the case of any $p$ such
that $0<p<0.5$. We first rewrite the transition probability $p_{i,j}$ of
Eq.~(\ref{eq:pij}) as
\begin{equation}
p_{i,j}=
(1-p)^n\left(\frac{1-p}{p}\right)^{-H_{j,k_i}},
\end{equation}
which for $0<p<0.5$ leaves it clear that $p_{i,j}$ decays exponentially with the
Hamming distance between $j$ and $k_i$ from the maximum value of $(1-p)^n$. This
maximum, as we have noted, is achieved for $j=k_i$, so evolving toward $i$'s
deterministic successor in a single time step is always exponentially more
likely than doing it toward any other CA state. Intuitively one might then
expect the occurrence of $H(\pi,\sigma)\approx 0$ to be commonplace, but we have
found this to be far from the truth. In fact, it all depends on the great
richness of detail we can always expect from CA behavior, particularly on how
the basins are laid out on the attractor field and whether the CA switches
basins in the event that some $j\neq k_i$ is picked when the current CA state is
$i$.

We proceed by singling out some rules for a more detailed discussion. Most of
these are highlighted in Tables~\ref{table:dist} and~\ref{table:msd} with a bold
typeface. We occasionally mention specific characteristics of a rule or its
basins, and for these the reader is referred to one of the available atlases
\cite{wl92,wa}.

First note that, though not commonplace, rules for which $H(\pi,\sigma)$ is
indistinguishable from $0$ within the six decimal places used in
Table~\ref{table:dist} do exist. These are class-1 rules 0 and 253; class-2
rules 15, 51, 204, and 240; and class-3 rules 60, 90 (the XOR rule), 105, and
150. For two of these rules, namely 0 and 253, the value of $H(\pi,\sigma)$ is
precisely $0$, since each of them gives rise to exactly one basin of attraction,
call it $B_1$, whence it follows that $\pi_{B_1}=\sigma_{B_1}=1$ no matter what
the stationary CA-state probabilities that make up $\pi_{B_1}$ turn out to be.
The value of $H(\pi,\sigma)$ is precisely $0$ also for four other rules, namely
15, 51, 204, and 240, but for an entirely different reason. What happens in
these cases is that the transition-probability matrix is doubly stochastic,
which as we have noted implies that the stationary distribution over the CA
states is uniform. For rules 51 and 204, in particular, double stochasticity is
a consequence of the matrices' being symmetric (i.e., $p_{i,j}=p_{j,i}$ for all
CA states $i$ and $j$). As for rules 60, 90, 105, and 150, $H(\pi,\sigma)$ is
probably only approximately equal to $0$, since the matrices do not seem to be
doubly stochastic.

Making the requirement on $H(\pi,\sigma)$ less stringent, for example by
replacing indistinguishability from $0$ with $H(\pi,\sigma)<0.1$, turns up
further rules: class-1 rules 249, 251, 252, and 254; class-2 rules 12, 26, 27,
29, 38, 205, 210, 220, 222, and 243; class-3 rules 30 and 45; and even one of
the elusive class-4 rules, namely rule 193 (more widely recognized through its
equivalent by both negation and reflection, the celebrated rule 110, known to be
capable of universal computation). The class-1 additions to the list are not
really surprising, since in all four cases nearly all CA states cluster into one
single basin and therefore our argument above for rules 0 and 253 essentially
continues to hold (though approximately). As for the remaining additions
(the class-2 through class-4 rules), no readily discernible characteristic seems
to stand out that might help explain the relative proximity of the two
distributions, not even inside each class.

Aside from these $27$ zero or near-zero cases of the Hellinger distance, the
remaining $61$ rules in Tables~\ref{table:dist} and~\ref{table:msd}, at least
for our small sample of $n$ and $p$ values, all give rise to stationary basin
probabilities that differ from those of the deterministic case (with initial CA
states chosen uniformly at random) to some substantial extent. Singling out some
rules on the higher extreme of distance values is not as clear-cut a task as
picking the zeroes. As we mentioned earlier, the theoretical maximum distance of
$1$ can never be achieved for distributions that are strictly positive
everywhere, so figuring out the actual maximum for elementary CA is far from a
trivial task.

What we do is then to highlight those rules that, across our small sample of $n$
and $p$ values, are on the far side of the (admittedly arbitrary) threshold
of $H(\pi,\sigma)=0.45$. Doing this yields four rules, all in class 2 and
italicized in the tables: rules 19, 23, 232, and 236. Once again it is hard to
discern any explaining characteristics, but from Table~\ref{table:msd} it is
clear that all four rules have in common the facts that $\rho_\mathrm{mean}$ is
substantially larger than $1$ (but less so as $p$ is increased) and that
$\rho_\mathrm{s.d.}$ is often smaller than $1$ (but growing as $p$ is
increased). That is, for small $p$ the distribution is more concentrated on
larger basins, all relative to the basin-size distribution arising from the
uniform distribution on CA states. This becomes less so as $p$ is increased and
the already discussed limit, as $p$ is driven toward $0.5$, is approached.

\section{Immunity as computation}\label{sec:disc2}

The present study has hinged on Eq.~(\ref{eq:disobey}), a simple probabilistic
expression of a cell's ability to alter its state differently than the CA rule
in use directs it to, at every time step and independently of all other cells.
If we view the CA states as states of the body, including the portion of it
known as the immune system, then the evolution of CA states in time stands not
only for the natural succession of body states but also for the computation of
such states by the immune system. Given this context, the adoption of the
spatially and temporally local probabilistic alterations to the CA rule given in
Eq.~(\ref{eq:disobey}) is an attempt to summarize several phenomena originating
from the uncertainty that is inherent to every biological process. Such
uncertainty drives adaptation, gives rise to diversity as well as disease, and
fuels the appearance of idiotypes never before seen in the body and with them
the possibility of better immunity through learning.

Though inherently stochastic, our model is also inherently dependent on a fixed
CA rule. This is clear already in Eq.~(\ref{eq:disobey}) itself, where we recall
that $b$ stands precisely for the cell's next state according to such a fixed
rule. Moreover, although Eq.~(\ref{eq:disobey}) makes every state update of
every cell nondeterministic, globally it is always exponentially more likely to
evolve to the CA state the rule mandates than to any other CA state. This means
that the clustering of CA states into basins, though no longer unbreachable, is
still meaningful and can be exploited as we adopt the modified CA as metaphors
of immunity as computation. For example, each basin can be viewed as
encompassing CA states that are equivalent from the perspective of the immune
system as it computes the state of the body. Some possibilities that come to
mind are basins representing a healthy or unhealthy body, others representing
a body under recovery through the action of the immune system, and still many
others as details are brought into the picture.

In such a setting, changes in the CA state other than those mandated by the
underlying CA rule can be interpreted in a variety of ways: e.g., inter-basin
transitions may stand for the appearance of or the recovery from diseases, as
well as to adaptation into a distinct, though still healthy, set of states;
intra-basin transitions, in turn, may represent change that nevertheless does
not fundamentally alter the state of the body as far as being healthy is
concerned. So far we have explored this landscape by simply asking what the
effects of Eq.~(\ref{eq:disobey}) might be in terms of fundamentally deviating
the CA from its traditional excursion into the field of attractor basins under
the CA rule in question. We have discovered CA rules in all four of Wolfram's
classes for which no fundamental deviation exists while still allowing the CA to
occasionally drift in and out of the field's basins. 

It is telling that we should find such behavior already in the simplest of CA,
viz.\ elementary CA, and already for the very small ones we investigate in this
work. Moving forward will require the investigation of more complex CA, at the
same time higher-dimensional, larger, and governed by larger-neighborhood rules.
We expect that these enriched scenarios will provide many useful possibilities
to characterize immunity as computation. In our view, the importance of
characterizations such as this can hardly be overstated: Even as we write,
immunotherapy is being hailed as a fundamental breakthrough in cancer treatment
(cf.\ \cite{c13}, as well as \cite{g13} and related content), and theoretical
modeling is bound to be instrumental in better understanding this and other
applications.

\section{Concluding remarks}\label{sec:concl}

An important characteristic of our model is its reliance on one single
parameter, the probability $p$. Assuming that it acts at each cell independently
of all others has allowed the transition probability $p_{i,j}$, from CA state
$i$ to CA state $j$ in a single step, to be written as in Eq.~(\ref{eq:pij}),
which in turn implies the ergodicity of the corresponding Markov chain whenever
$p>0$. The model is then conceptually simple, but studying it requires the
Markov chain's stationary probabilities to be found, which by virtue of the
model's inherent combinatorial growth in the general case quickly becomes
computationally burdensome if not downright intractable.

Further research should then first concentrate on looking for those CA rules, if
any exist, for which the transition matrix can somehow be simplified so that
some facilitating structure emerges. We already know that, for $p<0.5$, the
dominant probability on any of the matrix's rows, say the $i$th, is
$p_{i,k_i}=(1-p)^n$. Not only this, but $p_{i,j}$ for any $j\neq k_i$ is smaller
than $p_{i,k_i}$ by the exponentially decaying factor of
$[(1-p)/p]^{-H_{j,k_i}}$. The key to solving the Markov chain associated with
certain rules may lie precisely in ignoring such vanishingly small
probabilities, but to our knowledge substantial further research is needed to
ascertain this.

\subsection*{Acknowledgments}

We acknowledge partial support from CNPq, CAPES, a FAPERJ BBP grant, and the
joint PRONEX initiative of CNPq and FAPERJ under contract E-26/110.550/2010.

\bibliography{immca}
\bibliographystyle{plain}

\end{document}